\documentclass[11pt]{article}
\usepackage{amssymb}
\usepackage{graphics}
\usepackage{epsfig}
\usepackage{a4wide}
\usepackage{cite}
\usepackage{multirow}
\usepackage{color}

\newcommand\fverb{\setbox\fverbbox=\hbox\bgroup\verb}
\newcommand\fverbdo{\egroup\medskip\noindent%
            \fbox{\unhbox\fverbbox}\ }
\newcommand\fverbit{\egroup\item[\fbox{\unhbox\fverbbox}]}
\newcommand{\pphza}{$pp \to HZ\gamma+X$ }
\newbox\fverbbox

\begin{document}
%-------------------------------------------------------------------------------------------------------

\title{ $HZ\gamma$ production at $14~{\rm TeV}$ LHC in next-to-leading order QCD }

\author{ Xiong Shou-Jian$^a$, Ma Wen-Gan$^a$, Guo Lei$^b$, Zhang Ren-You$^a$, Chen Chong$^a$ and Song Mao$^c$ \\
{\small $^a$ Department of Modern Physics, University of Science and Technology of China, }  \\
{\small $~~$ Hefei, Anhui 230026, P.R.China} \\
{\small $^b$ Department of Physics, Chongqing University, Chongqing 401331, P.R. China} \\
{\small  $^c$ School of Physics and Material Science, Anhui University, Hefei, Anhui 230039, P.R.China} }

\maketitle \vskip 15mm
\begin{abstract}
We investigate the process $pp \to HZ\gamma+X$ at the $\sqrt{s}=14~{\rm TeV}$ LHC up to the QCD next-to-leading order (NLO), and discuss the kinematic distributions of final products after on-shell Higgs and $Z$-boson decays by adopting the narrow width approximation. The dependence of the leading order (LO) and the QCD NLO corrected integrated cross sections on the factorization/renormalization scale is studied. Our results show that the LO integrated cross section and kinematic distributions are significantly enhanced by the NLO QCD corrections, and the NLO QCD $K$-factor strongly depends on the observables and phase space. We conclude that in precision experimental data analyse for probing the $HZ\gamma$ coupling we should consider the NLO QCD corrections and put proper constraints on lepton-pair invariant mass to reduce the background.
\end{abstract}


\vfill \eject \baselineskip=0.32in

\renewcommand{\theequation}{\arabic{section}.\arabic{equation}}
\renewcommand{\thesection}{\arabic{section}}

\makeatletter      % '@' is now a normal "letter" for TeX
\@addtoreset{equation}{section}
%\@addtoreset{figure}{section}
%\@addtoreset{table}{section}
\makeatother       % '@' is restored as a "non-letter" character for TeX

\par
\section{Introduction}
\par
In the standard model (SM) the Higgs boson serves for the breaking of the electroweak symmetry and the generation of the fundamental particle masses \cite{sm, higgs}. Studying the Higgs mechanism is one of the main goals of the CERN Large Hadron Collider (LHC). In 2012 both the ATLAS and CMS collaborations announced the discovery of a new boson, whose properties are relatively close to the long awaited SM Higgs boson with mass of $m_H \sim 126~{\rm GeV}$ \cite{higgs1, higgs2}. After the discovery of the Higgs boson, the main task of further experiments is to determine its properties. Particularly, the precise determination of the Higgs boson couplings is imperative for verifying the validity of the SM and the existence of new physics at high energy scale.

\par
The LHC first runs at $7$ and $8~ {\rm TeV}$ are completed, and the latest data from them show that all the properties \cite{ATLASnew1,ATLASnew2,ATLASnew3,CMSnew1,CMSnew2} of the new boson measured so far are well consistent with that of the SM Higgs boson, but it is well known that there are some theoretical difficulties associated with the SM Higgs sector. For example, the famous hierarchy problem, which is associated with the quadratic radiative corrections to the SM Higgs mass, is one of the difficulties, and there is no way to solve this problem in the SM. So new physics effects are still expected to solve these difficulties. There are many model candidates of new physics predicting sizeable deviations from the the Higgs couplings in the SM. In the SM some Higgs couplings of the types $g_{HVV^{\prime}}$ and $g_{HVV^{\prime}V^{\prime\prime}}$ are absent at the tree-level, and they would be particularly sensitive to new physics \cite{loop-induced1,loop-induced2}. With the increases of the LHC luminosity and colliding energy, we can collect statistically enough events for most of the important multi-body production processes. Obviously, precision measurements require accurate theoretical predictions for both signal and background. In the last few years, the phenomenological results including the next-to-leading order (NLO) QCD corrections to the Higgs boson production associated with di-gauge-boson at the LHC, such as $pp \to HWW$, $pp \to HW^{\pm}\gamma$, $pp \to HW^{\pm}Z$, have been studied \cite{song2009,Mao:2013dxa,Liu:2013cla}.

\par
The $HZ\gamma$ production at the LHC also offers the possibility to directly investigate the $H Z \gamma$, $HZZ\gamma$ and $HZ\gamma\gamma$ anomalous Higgs gauge couplings \cite{HZgamma-anomaly,HVV@ee}, as they would cause deviations from the SM predictions. Moreover, an accurate estimate of the \pphza process followed with subsequent Higgs and $Z$-boson decays could provide the direct observable predictions in searching for possible new physics.
With the transverse momentum and rapidity cuts of $p_{T, cut}^{\gamma} = 5 \sim 20~ {\rm GeV}$ and $y_{cut}^{\gamma} = 2.5$ on the final photon, the cross section for the $HZ\gamma$ production at the $14~{\rm TeV}$ LHC is about $2.4 \sim 6.0~ fb$, and therefore only about $170 \sim 420~ fb^{-1}$ integrated luminosity is required to produce about 1000 events.

\par
In this paper, we make a precision calculation for the \pphza process at the LHC including the NLO QCD corrections with on-shell Higgs and $Z$-boson decays in the narrow width approximation (NWA), but do not provide the detailed strategy to extract the information of the anomalous Higgs gauge couplings from the $pp \rightarrow H Z \gamma + X$ process. In section II we give the description of the analytical calculations for the LO cross section and the NLO QCD radiative corrections to the \pphza process. In section III we present some numerical results and discussions. Finally, a short summary is given.

\vskip 5mm
\section{Description of the computation }
\par
In this section we describe the analytical calculations at the LO and QCD NLO for the $pp \to HZ\gamma+ X$ process.

\vskip 5mm
{\bf A. LO calculation }
\vskip 5mm
\par
\begin{figure}
\begin{center}
\includegraphics[scale=1.0]{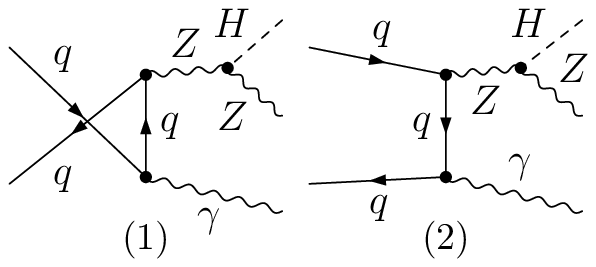}
\caption{ \label{fig1} The LO Feynman diagrams for the partonic process $q\bar{q} \to HZ\gamma$. }
\end{center}
\end{figure}

In our calculations we employ FeynArts 3.4 package \cite{fey} to generate LO and QCD NLO Feynman diagrams and their corresponding amplitudes. The algebraic manipulations on the amplitudes are implemented by applying FormCalc 5.4 programs \cite{formloop}. We neglect the masses of $u$-, $d$-, $c$-, $s$-quarks. Due to the smalless of (anti)bottom-quark density in proton, the LO contribution to the cross section from the $pp \to b\bar{b} \to HZ\gamma+X$ process at the $14~{\rm TeV}$ LHC is less than $0.6\%$. Therefore, we do not consider the partonic process of $b\bar{b}$ annihilation in our calculations. Then the contributions to the cross section for the parent process $pp \to HZ\gamma+X$ come from the following partonic processes,
\begin{equation}
\label{process1} q(p_1)+ \bar{q}(p_2) \to H(p_3)+Z(p_4)+\gamma(p_5),~~~~(q = u,d,c,s),
\end{equation}
where $p_{1}$, $p_{2}$ and $p_{3}$, $p_{4}$, $p_{5}$ represent the four-momenta of the incoming partons and the outgoing $H$, $Z$ and photon, respectively. The LO Feynman diagrams for the partonic process $q\bar{q} \to HZ\gamma$ are shown in Fig.\ref{fig1}. Since the Yukawa coupling strength is proportional to fermion mass and the $u$-, $d$-, $c$-, $s$-quarks are considered as massless, there is no contribution from the Feynman diagrams with internal Higgs boson line and the Higgs emission from initial quark. The LO amplitude for the $q\bar{q} \to HZ\gamma$ partonic process involves QED soft and collinear IR singualrities since the photon is radiated from massless quark. To avoid these QED IR singularities and obtain an IR-safe LO result, we take the transverse momentum and rapidity cuts on the final photon ($p_{T,cut}^{\gamma}$, $|y_{cut}^{\gamma}|$) as declared in the following section. The LO matrix element for the partonic process $q\bar{q} \rightarrow H Z \gamma$ can be expressed as
\begin{eqnarray}
\label{LO}{\cal M}_{LO} &=& {\cal M}^{u}_{LO} + {\cal M}^{t}_{LO} \nonumber\\
&=& \frac{i e^3Q_{q} m_{Z}}{ s^{2}_{w} c^{2}_{w}} \bar{v}(p_2)
\gamma^{\mu}(g^{q}_{V} -g^{q}_{A} \gamma^{5}) \frac{(\rlap/p_1 - \rlap/p_5)}{(p_1 - p_5)^2}
\gamma^{\nu}u(p_1)\frac{1}{(p_3+p_4)^2 - M^2_{Z}}
\epsilon^{*}_{\mu}(p_4)\epsilon^{*}_{\nu}(p_5) \nonumber\\
&+& \frac{i e^3Q_{q} m_{Z}}{s^{2}_{w} c^{2}_{w}}\bar{v}(p_2) \gamma^{\nu}
\frac{(\rlap/p_5 - \rlap/p_2)}{(p_5 - p_2)^2}
\gamma^{\mu}(g^{q}_{V} - g^{q}_{A}\gamma^{5})u(p_1)
\frac{1}{(p_3+p_4)^{2}-M^{2}_{Z}} \epsilon^{*}_{\mu}(p_4) \epsilon^{*}_{\nu}(p_5), \nonumber\\
\end{eqnarray}
where $s_w=\sin \theta_W$, $c_w=\cos \theta_W$, $g^{q}_{V}$ = $\frac{1}{2}T^{3}_{q} - Q_{q}\sin^{2}\theta_{W}$, $g^{q}_{A} = \frac{1}{2}T^{3}_{q}$, $T^{3}_{q}$ and $Q_{q}$ are the third component of weak isospin and the electric charge of quark $q$ separately.
%%%%%%%%%%%%%%%%%figure%%%%%%%%%%%%%%%%%%%%%%%%%%%%%%

\par
The LO cross section for the partonic process $q \bar{q} \to H Z \gamma$ can be obtained by performing the integration over the phase space expressed as below,
\begin{eqnarray}
\hat{\sigma}^0_{q\bar{q}} =  \frac{(2\pi)^4}{4 |\vec{p}| \sqrt{\hat{s}}} \int \overline{\sum}  \left|  {\cal M}_{LO}^{q\bar{q}}  \right|^2  d \Omega_3,
\end{eqnarray}
where $\vec{p}$ is the three-momentum of one initial parton in the center-of-mass system (c.m.s), $\sqrt{\hat{s}}$ is the colliding energy in partonic c.m.s, the summation is taken over the spins and colors of the initial and final states, and the bar over the summation indicates the averaging over the intrinsic degrees of freedom of initial partons. $d\Omega_3$ is the three-body phase space element defined as
\begin{eqnarray}
d\Omega_3=\delta^{(4)} \left( p_1+p_2-\sum_{i=3}^5 p_i \right)
\prod_{j=3}^5 \frac{d^3 \vec{p}_j}{(2 \pi)^3 2 E_j}.
\end{eqnarray}

By convoluting $\hat{\sigma}^0_{q\bar{q}}$ with the parton distribution
functions (PDFs) of the colliding protons, we obtain the LO total cross section for the parent process
$pp \to HZ\gamma + X$ as
\begin{eqnarray}
\sigma_{LO} =\sum_{q} \int_0^1 dx_1 dx_2 \biggl[ G_{q/P_1}(x_1, \mu_f) G_{\bar{q}/P_2}(x_2, \mu_f) \hat{\sigma}^0_{q\bar{q}}(\sqrt{\hat{s}} = x_1 x_2 \sqrt{s}) + (1 \leftrightarrow 2) \biggr],
\end{eqnarray}
where $G_{q/P}$ represents the PDF of parton $q$ in proton $P$, $x_i~(i=1,2)$ describes the momentum fraction of a parton in proton, $\sqrt{s}$ is the colliding energy in the rest frame of proton-proton system, and $\mu_f$ is the
factorization scale.

\vskip 5mm
{\bf B. NLO calculation }
\vskip 5mm
\par
\begin{figure}
\begin{center}
\includegraphics[scale=0.8]{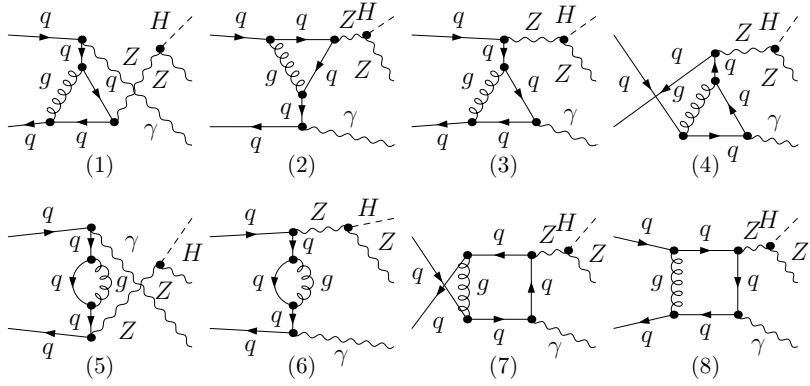}
\caption{ \label{fig2} The one-loop Feynman diagrams for the partonic process $q\bar{q} \to HZ\gamma$. }
\end{center}
\end{figure}

In the NLO calculations we use the dimensional regularization (DR) method in $D=4-2 \epsilon$ dimensions to regularize the ultraviolet (UV) and infrared (IR) divergences. The NLO QCD corrections to the \pphza process are constituted distinctly by the following three parts: (1) the virtual correction, (2) the real gluon and light-(anti)quark emission corrections, (3) the collinear counterterms of the PDFs. The virtual NLO QCD correction to the $q\bar{q} \to HZ\gamma$ partonic process consists of self-energy, vertex, box and counterterm diagrams. The one-loop Feynman diagrams are shown in Fig.\ref{fig2}. We follow the definitions of tensor and scalar one-loop integral functions in Refs.\cite{Passarino,denner2}, and use the Passarino-Veltman (PV) method \cite{Passarino,denner3} to reduce tensor integrals to the linear combinations of tensor structures and coefficients, where the tensor structures depend on the external momenta and the metric tensor, while the coefficients depend on scalar integrals and kinematic invariants. The whole reduction manipulations of a tensor integral to the lower-rank tensors and further to scalar integrals, is done and numerically calculated by using the LoopTools-2.2 library \cite{formloop} and the FF package \cite{ff}.
%%%%%%%figure%%%%%%

\par
In the virtual correction calculation we need the wave function renormalization constants for quark fields. We introduce the renormalization constants $\delta Z_{\psi_{q,L,R}}$ for massless quark ($q=u,d,c,s$) fields defined as
\begin{eqnarray}
\psi^{0}_{q,L,R} = (1+\delta Z_{\psi_{q,L,R}})^{1/2}\psi_{q,L,R}.
\end{eqnarray}
In the modified minimal subtraction ($\overline{MS}$) renormalization scheme the renormalization constants for the massless quarks are expressed as
\begin{eqnarray}
\delta Z_{\psi_{q,L}} &=&\delta Z_{\psi_{q,R}} = -\frac{\alpha_{s}}{4\pi}C_{F}(\Delta_{UV}-\Delta_{IR}),
\end{eqnarray}
where $C_F=4/3$, $\Delta_{UV}=\frac{1}{\epsilon_{UV}}-\gamma_E + \ln (4\pi)$ and $\Delta_{IR}=\frac{1}{\epsilon_{IR}}-\gamma_E + \ln (4\pi)$.

\par
After the reduction for tensor integrals, the amplitude for loop corrections involving one-loop scalar integrals contains both UV and IR divergences. The UV divergence is vanished after performing the renormalization procedure. But the total QCD NLO amplitude for the subprocess $q\bar{q} \to HZ\gamma$ still contains QCD soft/collinear IR singularities. We adopt the expressions in Ref.\cite{IRDV} to deal with the QCD IR divergences in Feynman integrals, and apply the expressions in Refs.\cite{OneTwoThree,Four,Five} to implement the numerical evaluations for the QCD IR-finite parts of $N$-point scalar integrals. According to the Kinoshita-Lee-Nauenberg (KLN) theorem \cite{KLN}, these IR singularities will be cancelled by adding the contributions of the real gluon/light-(anti)quark emission subprocesses, and redefining the PDFs at the QCD NLO.

\par
\begin{figure}
\begin{center}
\includegraphics[scale=0.8]{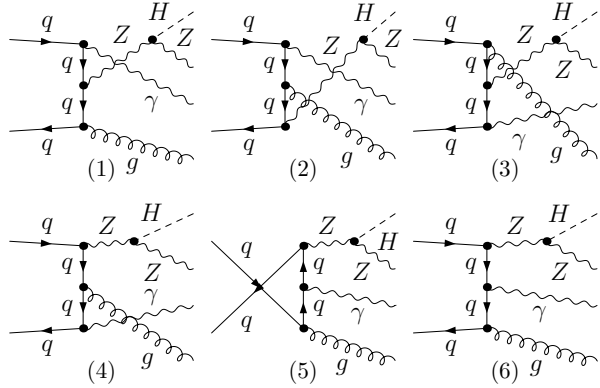}
\caption{ \label{fig3} The Feynman diagrams for the real gluon emission subprocess $q \bar{q} \to HZ \gamma + g$. }
\end{center}
\end{figure}
\begin{figure}
\begin{center}
\includegraphics[scale=0.8]{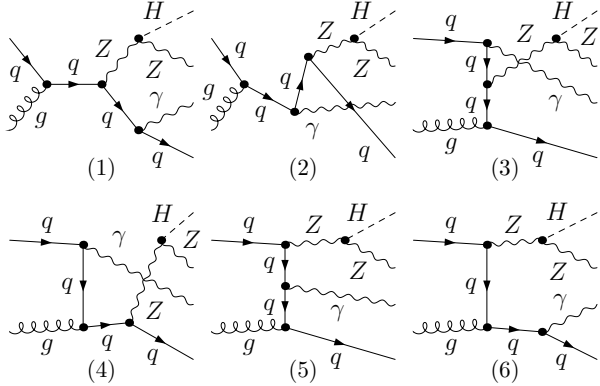}
\caption{ \label{fig4} The Feynman diagrams for the real light-quark emission subprocess $qg \to HZ \gamma + q$. }
\end{center}
\end{figure}

Since we put the transverse momentum, rapidity cuts on the final photon and a resolution cut on the photon and final jet throughout our LO and QCD NLO calculations, the numerical results for the real gluon/light-(anti)quark emission subprocesses are QED IR safe. While the real gluon/light-quark emission processes contain the QCD soft and collinear IR singularities. Technically, we isolate the QCD soft and collinear IR singularities by adopting the two cutoff phase space slicing (TCPSS) method \cite{TCPSS}. The Feynman diagrams for the real gluon and light-quark emission subprocesses are shown in Fig.\ref{fig3} and Fig.\ref{fig4}, separately. Before our numerical calculations, we checked the UV and IR divergence cancelations both analytically and numerically. To verify the implementation of the TCPSS in right way, the independence of the NLO QCD corrected total cross section on the soft cutoff $\delta_{s}$ are checked in the range of $ 1 \times 10^{-5} < \delta_{s} < 1 \times 10^{-3}$ with $\delta_{c} = \delta_{s}/50$. In the further numerical calculations, we fix $\delta_{s} = 1 \times 10^{-3}$ and $\delta_{c} = 2 \times 10^{-5}$. Furthermore, we numerically compared our NLO QCD corrected cross sections with those obtained by using program MadGraph5$\_$aMC@NLO of version 2.2.2 \cite{MG5NLO}, and find they are in good agreement with each other within the Monte Carlo errors.
%%figure

\vskip 5mm
\section{Results and discussions}
\par
In this section we present and discuss the numerical results for the $H Z \gamma$ associated production at the $\sqrt{s} = 14~{\rm TeV}$ LHC at both the LO and QCD NLO. We use the CTEQ6L1 and CT10nlo PDFs in the LO and NLO calculations, respectively. The strong coupling constant is determined by taking one-loop and two-loop running $\alpha_{s}(\mu)$ for the LO and NLO calculations separately, and setting the QCD
parameter as $N_f=5$, $\Lambda_5^{LO} = 165~{\rm MeV}$ for the CTEQ6L1 and $\Lambda_5^{\overline{MS}} = 226~{\rm MeV}$ for the CT10nlo. We set the factorization and renormalization scales to be equal, and take $\mu=\mu_f = \mu_r = \mu_0$ by default unless stated otherwise. The central scale is defined as $\mu_0=E_T/2=\frac{1}{2} \sum\limits_i E_{T,i}$, where $E_{T,i}=\sqrt{p_{T,i}^2+m_i^2}$ and the summation is taken over all the transverse energies of final particles. The related SM input parameters are taken as \cite{PDG2012,higgs1, higgs2}
\begin{equation}
\begin{array}{llll}  \label{input1}
\alpha_{ew}^{-1}=137.036, &m_W=80.385~{\rm GeV},    &m_Z=91.1876~{\rm GeV},   &m_H=126~{\rm GeV}.
\end{array}
\end{equation}

\par
To strip the QED soft and collinear IR singularities at both the LO and QCD NLO, we put the following transverse momentum, rapidity and resolution cuts on the final photon, i.e.,
\begin{eqnarray} \label{isol-1}
&& p_{T}^{\gamma}> p_{T,cut}^{\gamma}, ~~ |y_{\gamma}| \le |y_{cut}^{\gamma}|=2.5,   \\
\label{isol-2}
&& R_{\gamma j}> \delta_0 ~~~~~{\rm or} ~~~~~p_{T}^{j}\leq p_{T}^{\gamma}\frac{1-\cos R_{\gamma j}}{1-\cos\delta_0}.
\end{eqnarray}
where $\delta_0$ is a fixed separation parameter which is set to be $0.7$. The condition of Eq.(\ref{isol-2}) implies that the final jet can arbitrarily close to the photon as long as the jet is soft enough. In this way, we can preserve the full QCD singularities, which cancels against the virtual part, but it does not introduce divergence from the interaction between photon and massless quark-jet \cite{Frixione98}. The limitation in Eq.(\ref{isol-2}) is to remove the QED collinear IR singularity due to a photon radiated from a final light-quark-jet $j$ in the NLO calculation for the real light-quark emission processes. Then we accept the $HZ\gamma+jet$ events only if all the limitations in (\ref{isol-1}) and (\ref{isol-2}) are satisfied.

\par
\begin{figure}
\begin{center}
\includegraphics[scale=0.3]{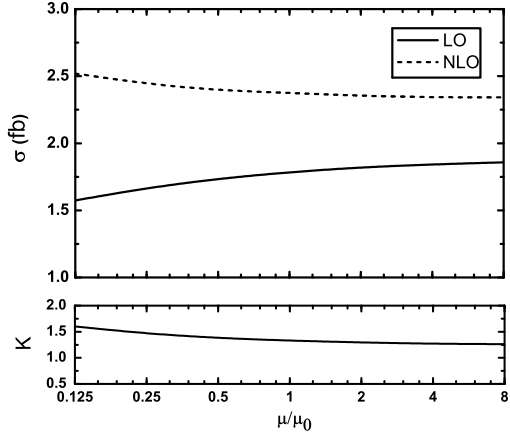}
\caption{ \label{fig5}  The dependence of the LO, NLO QCD corrected total cross sections and the corresponding $K$-factor for the \pphza process on the factorization/renormalization scale. Here we assume $\mu=\mu_r=\mu_f$ and define the central scale as $\mu_0 =E_T/2$. }
\end{center}
\end{figure}

In Fig.\ref{fig5} we display the renormalization/factorization scale dependence of the LO, NLO QCD corrected total cross sections and the corresponding $K$-factor for the \pphza process at the $\sqrt{s} = 14~{\rm TeV}$ LHC by setting $\mu_r=\mu_f=\mu$. The LO and NLO QCD corrected integrated cross sections are $1.785~fb$ and $2.37~fb$, respectively, and the corresponding $K$-factor is $1.33$ at the central scale $\mu=\mu_0$. If we defined the relative scale uncertainty as $\eta= \left[max(\sigma(\mu))-min(\sigma(\mu))\right]/\sigma(\mu_0)$ with $\mu \in [0.25\mu_0,~4\mu_0]$, we get $\eta = 9.85 \%$ and $4.24 \%$ for the LO and NLO QCD corrected corss sections, respectively. We can see that the dependence of the NLO QCD corrected total cross section on the factorization/renormalization scale is significantly reduced compared with that of the LO integrated cross section. This makes the theoretical predictions much more reliable.
%%figure
%%%%%%%%%%%%%%%%% Table $$$$$$$$$$$$$$$$$$$$$$$$$$$$$$$$$$$
\begin{table}[t]
\begin{center}
\begin{tabular}{c|c|c|c}
\hline \hline
  $p_{T,cut}^{\gamma}$ ($GeV$) & $\sigma_{LO}$ ($fb$) & $\sigma_{NLO}$ ($fb$) & $K$ \\
\hline \hline
  5 & 4.780(4) & 5.99(3) & 1.25\\
  10 & 3.105(3) & 4.01(2) & 1.29\\
  15 & 2.287(2) & 3.00(1) & 1.31 \\
  20 & 1.785(1) &  2.37(1)& 1.33 \\
\hline
\end{tabular}
\end{center}
\begin{center}
\caption{\label{tab1} The LO, NLO QCD corrected integrated cross sections and the corresponding $K$-factors with different cuts on $p_{T}^{\gamma}$ for the $HZ{\gamma}$ production at the $14~{\rm TeV}$ LHC.  }
\end{center}
\end{table}

\par
We present the LO, NLO QCD corrected integrated cross sections and the corresponding $K$-factors for different cuts on $p_{T}^{\gamma}$ in Tab.\ref{tab1}. We can see that the LO and NLO QCD corrected total cross sections strongly depend on $p_{T,cut}^{\gamma}$, while the $K$-factor behaves not so sensitively to the cut. As shown in Tab.\ref{tab1}, we can get a sizeable decrease in the total cross section with the increase of $p_{T,cut}^{\gamma}$ for the $pp \to HZ \gamma+X$ process. In following numerical calculations we fix $p_{T,cut}^{\gamma} =20~{\rm GeV}$ as the default choice.

\par
\begin{figure}
\centering
\includegraphics[scale=0.25]{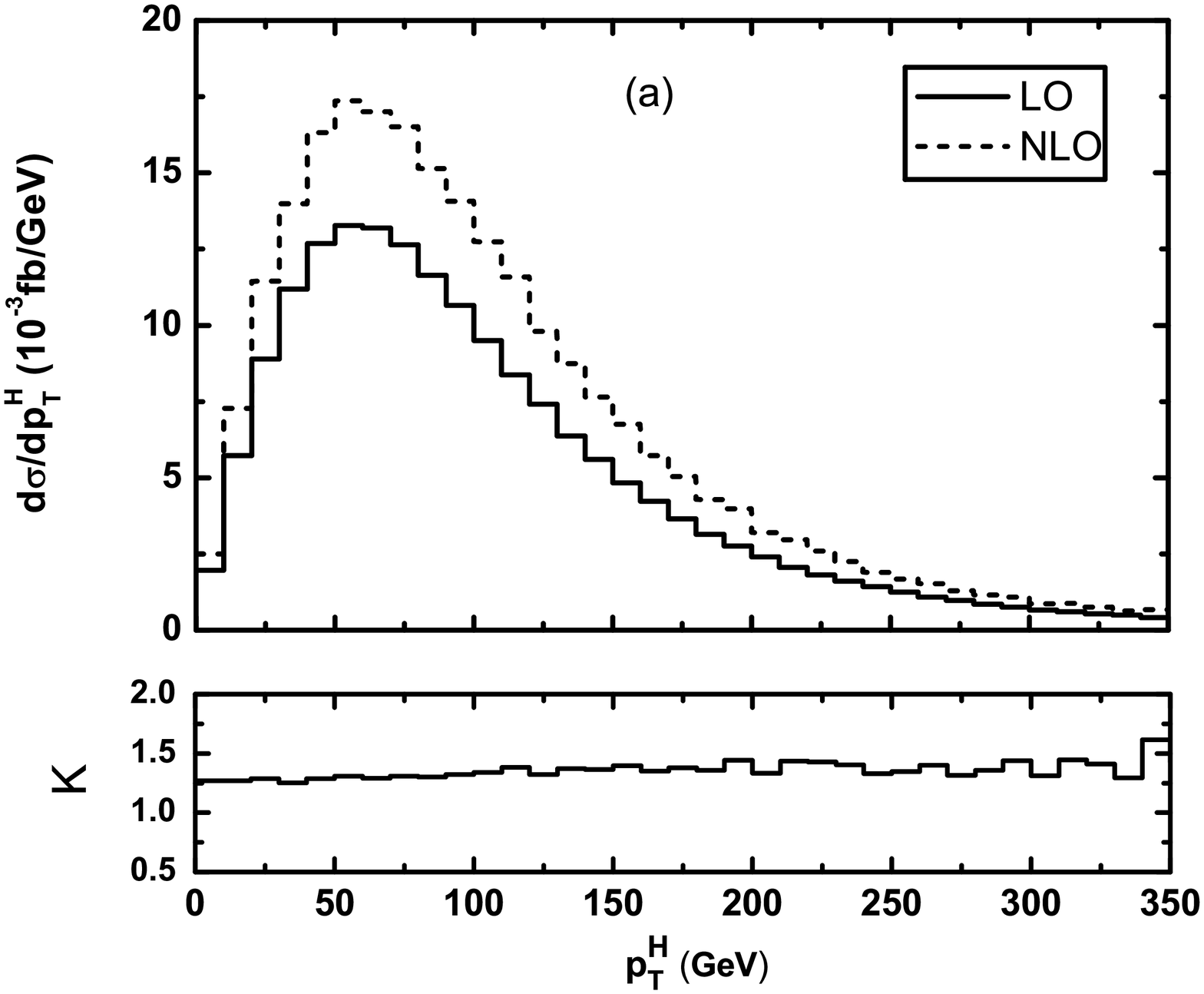}
\includegraphics[scale=0.25]{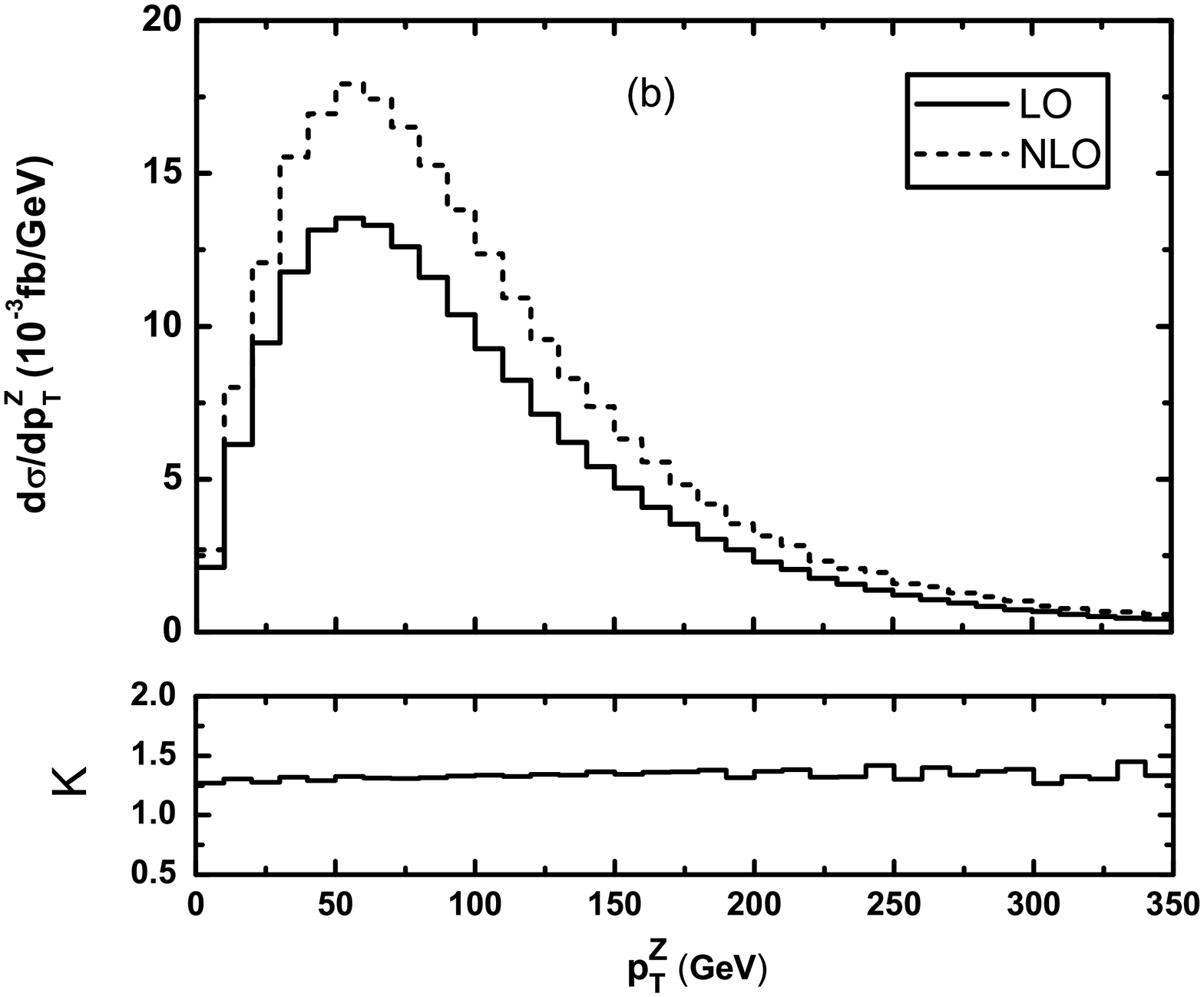}
\includegraphics[scale=0.25]{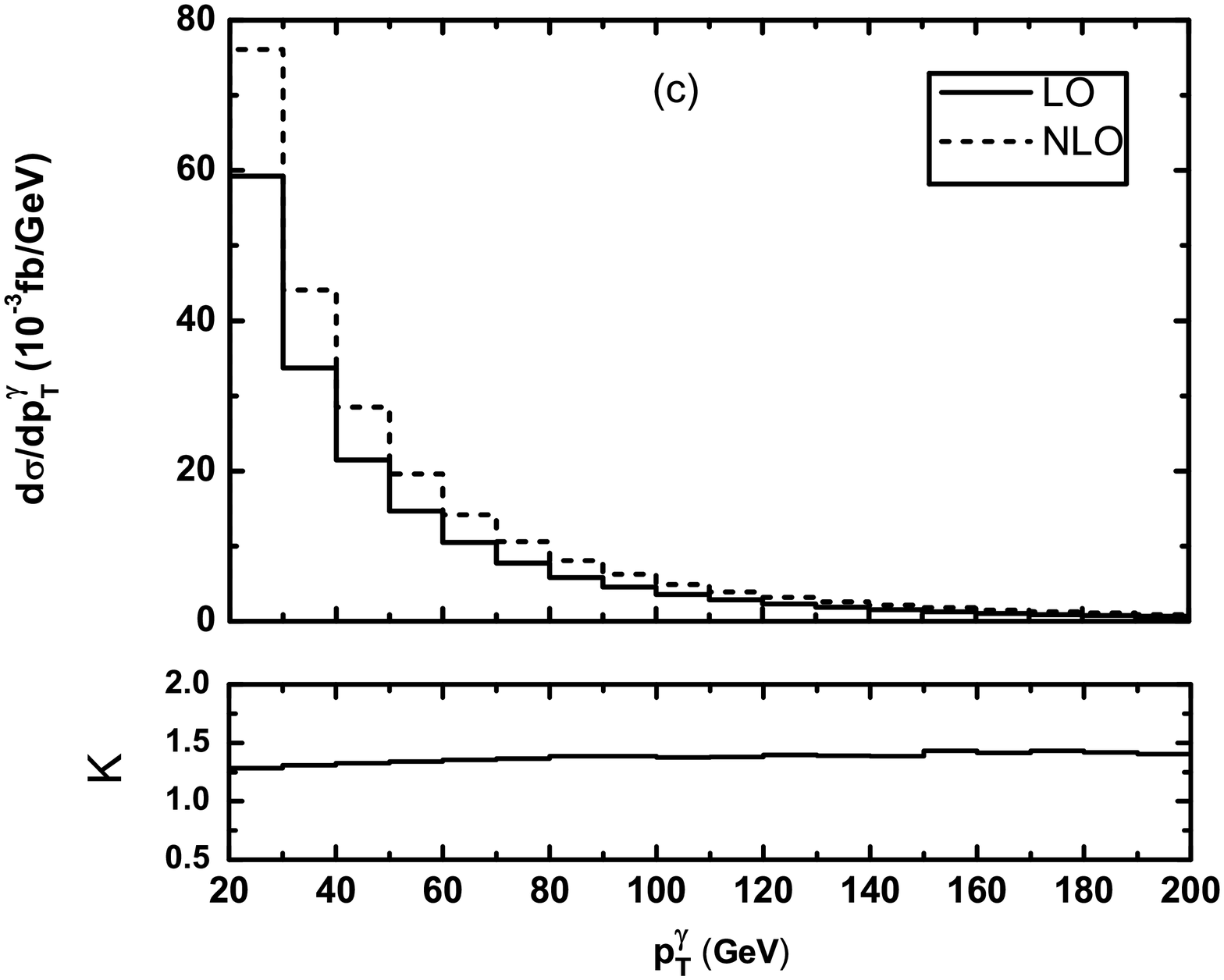}
\caption{\label{fig6} The LO, NLO QCD corrected transverse momentum distributions of the final particles and the corresponding $K$-factors for the \pphza process at the LHC. (a) $p_T^{H}$ distributions, (b) $p_T^{Z}$ distributions, (c) $p_T^{\gamma}$ distributions. }
\end{figure}

We depict the LO, NLO QCD corrected distributions of the transverse momenta and the corresponding $K$-factors for the Higgs, $Z$-boson and photon produced by the \pphza process at the $\sqrt{s} = 14~{\rm TeV}$ LHC in Figs.\ref{fig6}(a), (b) and (c), respectively. In Figs.\ref{fig6}(a) and (b), the curves for $\frac{d\sigma}{dp_T^{H}}$ and $\frac{d\sigma}{dp_T^{Z}}$ at the LO and QCD NLO peak at the position of $p_T \sim 50~{\rm GeV}$, and their $K$-factors are $1.31$ and $1.33$, respectively. Fig.\ref{fig6}(c) shows that both the LO and NLO QCD corrected $p_T^{\gamma}$ distributions decrease rapidly with the increase of the transverse momentum of photon. We can see from Figs.\ref{fig6}(a,b,c) that the transverse momentum distributions for the Higgs, $Z$-boson and photon ($d\sigma_{LO}/dp_T^{H}$, $d\sigma_{LO}/dp_T^{Z}$, $d\sigma_{LO}/dp_T^{\gamma}$) are significantly enhanced by the NLO QCD corrections.
%%%%figure%%%%%%
%%%%%%%%%%%%%%%%%%%%%%%%%%%%%%%%%%%%%%%%%%%%%%%%%%%%%%%%%%%%%%%%%

\par
The final photon can be directly detected in experiment, while the produced on-shell Higgs and $Z$-boson are unstable particles and can be detected by their decay products. In order to investigate the kinematic distributions of final directly detected particles, we apply the NWA in analysing the differential cross sections of the final Higgs and $Z$-boson decay products. We choose the Higgs boson decay channel of $H \to \tau^{+} \tau^{-}$ with $m_{\tau} = 1776.82~{\rm MeV}$ \cite{PDG2012} and $Z$-boson decay channel of $Z \to \ell^{+} \ell^{-}$ $(\ell = e, \mu)$ as the Higgs and $Z$-boson signals separately. By adopting HDECAY program \cite{Hdecay} with input parameters from Ref.\cite{PDG2012}, we get $Br(H \to \tau^{+} \tau^{-})=5.897\%$, and take $Br(Z \to \ell^{+} \ell^{-})=$ $Br(Z \to e^{+} e^{-})$ + $Br(Z \to \mu^{+} \mu^{-}) =$ $3.363\% + 3.366\%= 6.729\%$ \cite{PDG2012}. Then the signature for the $HZ\gamma$ production including the subsequent decays at the LHC can be written as
\begin{eqnarray}\label{channel}
pp \to HZ\gamma \to \tau^{+} \tau^{-} \ell^{+} \ell^{-}\gamma +X~~~  (\ell = e,\mu).
\end{eqnarray}
This signal is detected as an event including one $\tau$-pair, one $e(\mu)$-pair and a photon. For photon separation with other final particles in the rapidity-azimuthal angle plane, we impose the $R$ cuts between photon and other final particles as below:
\begin{equation}
\begin{array}{llll} \label{photoncuts}
R_{\tau \gamma} > 0.4,~ &R_{l\gamma} > 0.4, ~&R_{j\gamma} > 0.7,
\end{array}
\end{equation}
where $\ell = e,\mu$ and $j$ denotes a jet with transverse momentum $p_T^j > 30~{\rm GeV}$.

\par
The final photon in the signal process $pp \to H Z \gamma \to \tau^{+} \tau^{-} \ell^{+} \ell^{-} \gamma~(\ell = e,\mu)$ is only emitted from initial parton. However, the $HZ$ associated production followed by the subsequent decays of $H \to \tau^{+} \tau^{-} \gamma$, $Z \to \ell^{+}\ell^{-}$ or $H \to \tau^{+} \tau^{-}$, $Z \to \ell^{+}\ell^{-}\gamma$ also leads to the $\tau^{+} \tau^{-} \ell^{+} \ell^{-} \gamma$ event and contributes at the same order as the signal process. This process, denoted as
\begin{eqnarray}\label{background}
pp \to HZ \to \tau^{+} \tau^{-} \ell^{+} \ell^{-}\gamma +X~~~  (\ell = e,\mu),
\end{eqnarray}
is the main background in measuring the $HZ\gamma$ coupling via the $pp \to H Z \gamma \to \tau^{+} \tau^{-} \ell^{+} \ell^{-} \gamma + X$ process. We estimate the background process $pp \to H Z \to \tau^{+} \tau^{-} \ell^{+} \ell^{-} \gamma$ at the LO in the NWA by adopting CTEQ6L1 PDF, and take $\Gamma_{Z} = 2.4952~{\rm GeV}$ \cite{PDG2012}. The total decay width of SM Higgs boson is obtained by using HDECAY program as $\Gamma_{H}= 4.38 \times 10^{-3} ~{\rm GeV}$. In analysing the $\tau^{+} \tau^{-} \ell^{+} \ell^{-} \gamma$ events, we adopt the event selection criteria shown in Eqs.(\ref{isol-1}) and (\ref{photoncuts}), and impose the following invariant mass constraints on the final lepton pairs to suppress the background contribution:
\newpage
\begin{eqnarray}
&& m_H - \Delta < M_{\tau^+\tau^-} < m_H + \Delta, \nonumber \\
&& m_Z - \Delta < M_{\ell^+\ell^-} < m_Z + \Delta,~~~(\ell = e, \mu),~~~~~(\Delta = 5~ {\rm or}~ 10~ {\rm GeV}).
\end{eqnarray}
Then we obtain the background over signal as
\begin{eqnarray}
\frac{\sigma_{LO}(pp \to H Z \to \tau^{+} \tau^{-} \ell^{+} \ell^{-} \gamma)}{\sigma_{LO}(pp \to H Z \gamma \to \tau^{+} \tau^{-} \ell^{+} \ell^{-} \gamma)}
=
\left\{
\begin{array}{l}
8.1\%,~~~ \Delta = 10~{\rm GeV} \\
1.5\%,~~~ \Delta = 5~~{\rm GeV}
\end{array}
\right..
\end{eqnarray}
Therefore, we can conclude that the background events with photon radiated from final charged leptons can be reduced distinctly in probing the $HZ\gamma$ coupling by taking proper invariant mass constraints on $\tau$-pair and $e(\mu)$-pair. In the following, we neglect the background contribution and only consider the signal process $pp \to H Z \gamma \to \tau^{+} \tau^{-} \ell^{+} \ell^{-} \gamma$ in investigating the kinematic distributions of the final produced leptons.

\par
\begin{figure}[htbp]
\begin{center}
\includegraphics[scale=0.25]{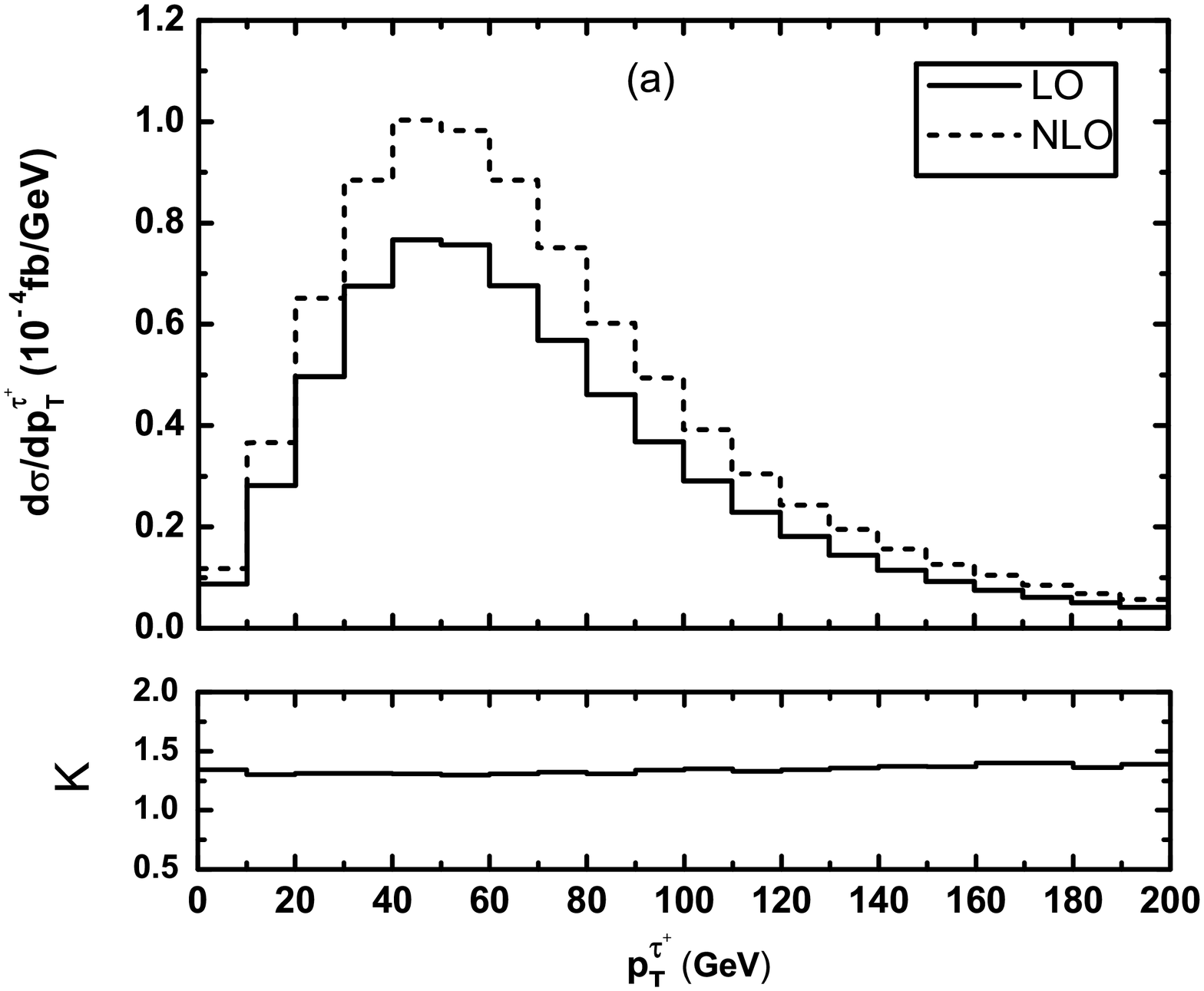}
\includegraphics[scale=0.25]{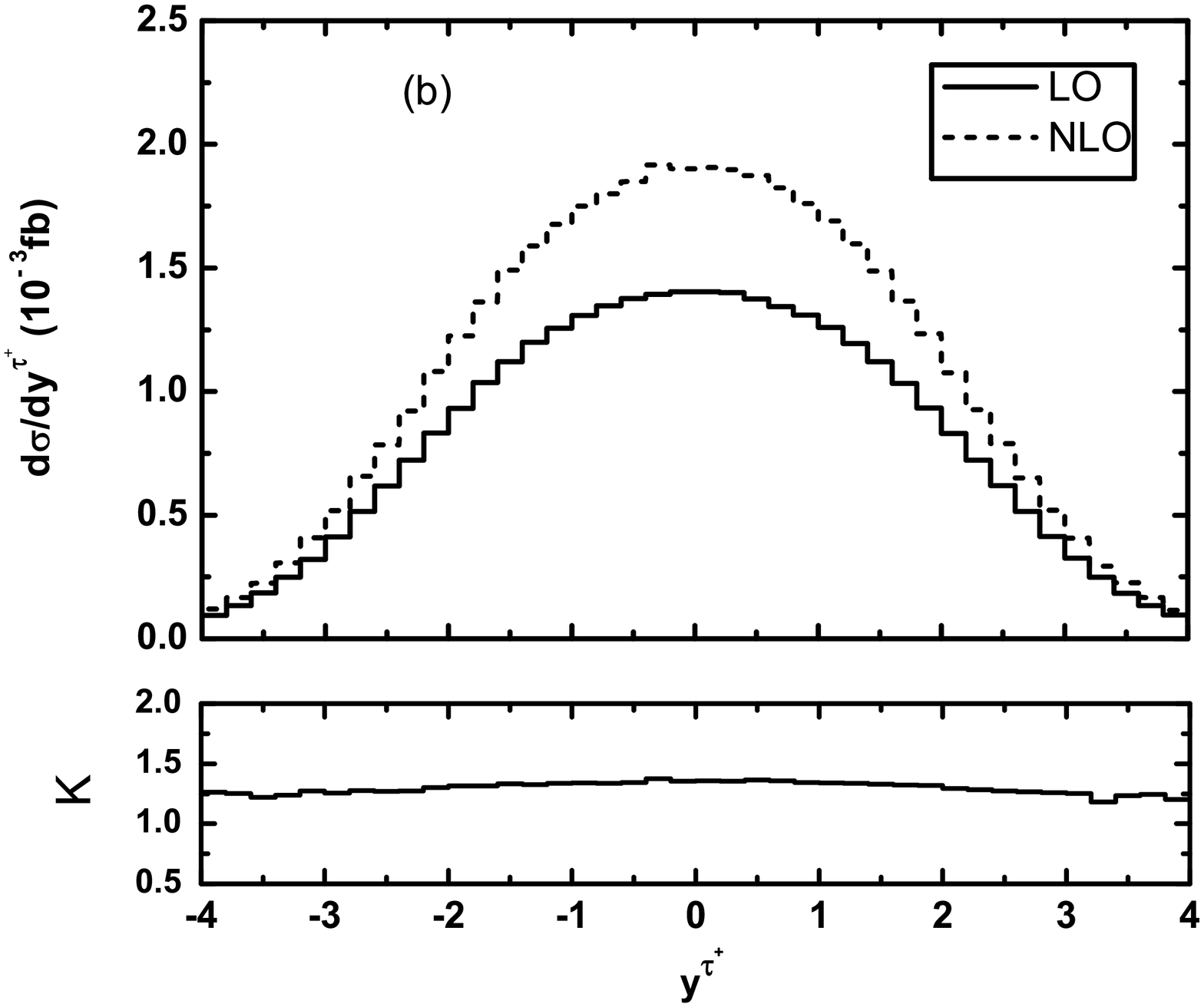}
\\~~  \\
\includegraphics[scale=0.25]{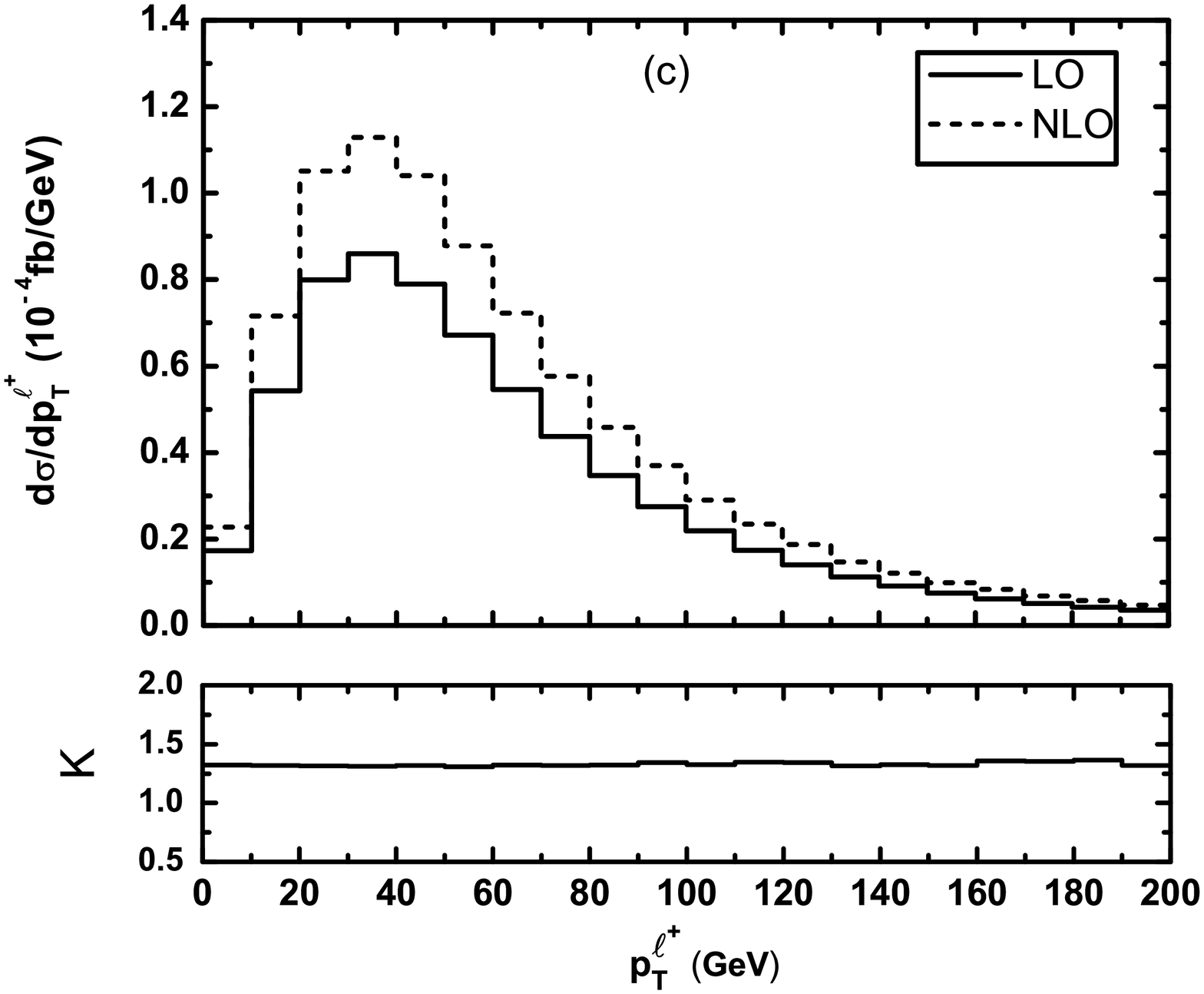}
\includegraphics[scale=0.25]{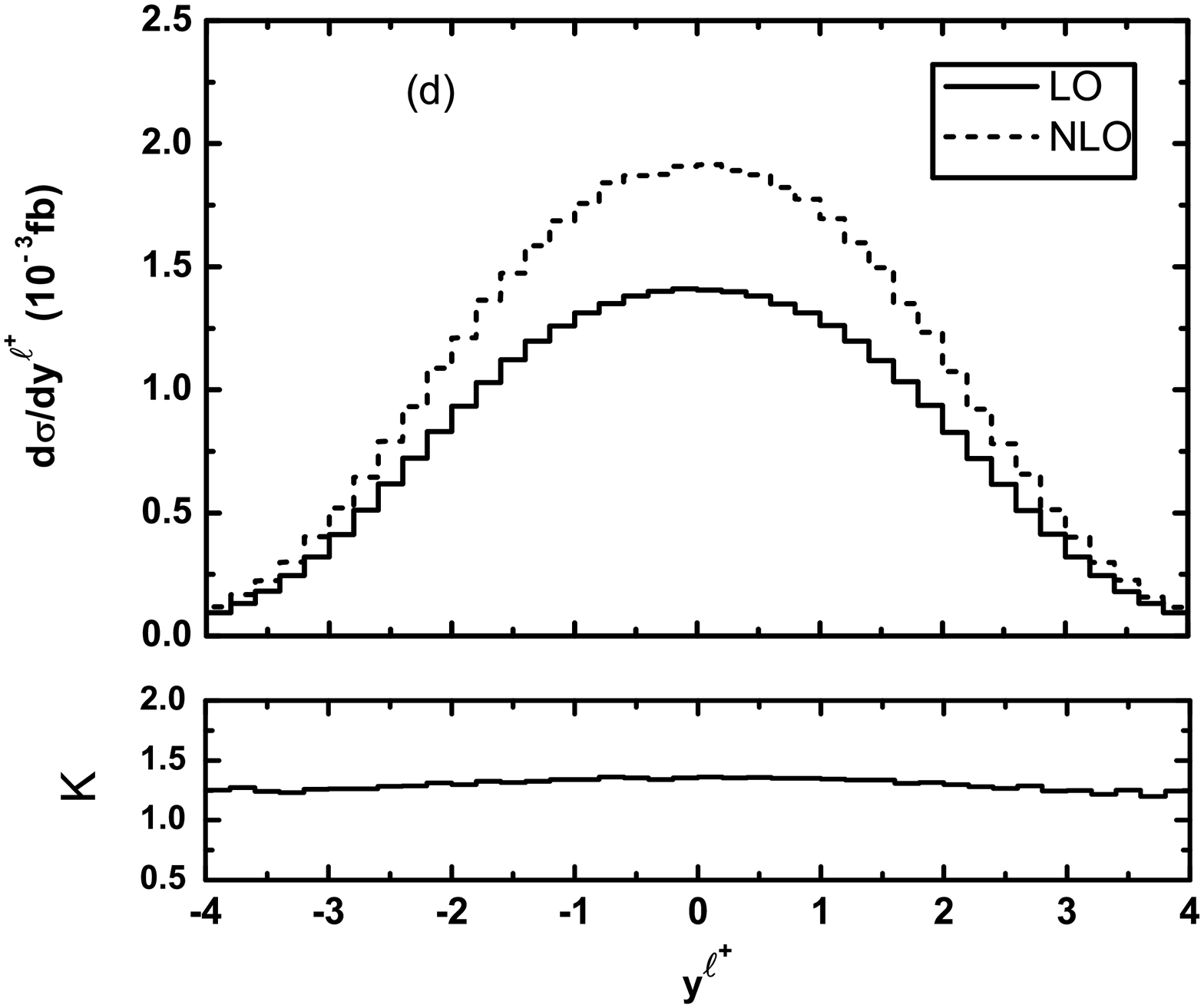}
\caption{\label{fig7} The LO, NLO QCD corrected transverse momentum and rapidity distributions of $\tau^{+}$ and the positively charged lepton, and corresponding $K$-factors for the $pp \to HZ\gamma+X \to \tau^{+} \tau^{-} \ell^+\ell^-\gamma+X$ processes at the $\sqrt{s}=14~{\rm TeV}$ LHC.  (a) $p_T^{\tau^{+}}$ distributions, (b) $y^{\tau^{+}}$ distributions, (c) $p_T^{\ell^+}~(\ell = e,\mu)$ distributions, (d) $y^{\ell^+}~(\ell = e,\mu)$ distributions. }
\end{center}
\end{figure}

In Figs.\ref{fig7}(a) and (c) we present the LO, NLO QCD corrected transverse momentum distributions of $\tau^{+}$ and positively charged lepton $\ell^+$ ($\ell = e, \mu$), and the corresponding $K$-factors at the $\sqrt{s}=14~{\rm TeV}$ LHC, respectively. As shown in Figs.\ref{fig7}(a) and (c), the QCD corrections always enhance the LO differential cross sections $d\sigma_{LO}/dp_T^{\tau^{+}}$ and $d\sigma_{LO}/dp_T^{\ell^+}$. Both the LO and NLO QCD corrected transverse momentum distributions of $\tau^{+}$ and $\ell^+$ reach their maxima at the positions of $p_T^{\tau^{+}}\sim 40~{\rm GeV}$ with $K = 1.31$ and $p_T^{\ell^+}\sim 30~{\rm GeV}$ with $K=1.31$, respectively. Figs.\ref{fig7}(b) and (d) are for the LO and NLO QCD corrected rapidity distributions of $\tau^{+}$ and $\ell^{+}$ separately. Both $y^{\tau^{+}}$ and $y^{\ell^{+}}$ reach their maxima at $y= 0$, with their $K$-factors being around $1.36$. We can see from all these four figures that the NLO QCD corrections do not change the line-shapes of the transverse momentum and rapidity distributions, while enhance the LO differential cross sections significantly in all the plotted kinematic regions. With the transverse momentum and rapidity cuts of $p_{T, cut}^{\gamma} = 5 \sim 20~ {\rm GeV}$ and $y_{cut}^{\gamma} = 2.5$ on the final photon, the accumulated luminosity of $1000 \sim 2660~ fb^{-1}$ is required to produce 25 $\tau^{+} \tau^{-} \ell^{+} \ell^{-} \gamma$ events via $pp \rightarrow H Z \gamma \rightarrow \tau^{+} \tau^{-} \ell^{+} \ell^{-} \gamma + X$ channel at the $14~ {\rm TeV}$ LHC.

%%%%%%%%%%%%%%%Figure%%%%%%%%%%%%%%%%%%%%%%%%%%%%%%

\vskip 5mm
\section{Summary}
\par
In this paper we investigate the NLO QCD corrections to the $HZ\gamma$ production followed by subsequent Higgs and $Z$-boson decays at the $\sqrt{s} = 14~{\rm TeV}$ LHC. We study the dependence of the LO and NLO QCD corrected cross sections on the factorization/renormalization scale, and our results show that the scale uncertainty of the NLO QCD corrected cross section is reduced compared with that of the LO cross section. We present the LO and NLO QCD corrected distributions of transverse momenta and rapidities of the decay products of Higgs and $Z$-boson. We find that the NLO QCD radiative corrections are significant, and notably modify the LO kinematic distributions. We see also that the $K$-factor is distinctly related to phase space region and kinematic observable. We conclude that the NLO QCD corrections should be considered in precision experimental data analyse in measuring the \pphza process, and the background events with a photon radiated from final charged lepton can be reduced in probing the $HZ\gamma$ coupling by putting proper invariant mass constraints on final $\tau$-pair and $e(\mu)$-pair.

\par
\section{Acknowledgments}
This work was supported in part by the National Natural Science Foundation of China (Grants. No.11275190, No.11375008, No.11375171).


\begin{thebibliography}{99}
\bibitem{sm}
S. L. Glashow, Nucl. Phys. {\bf 22}, 579 (1961); S. Weinberg, Phys.
Rev. Lett. {\bf 19}, 1264 (1967); A. Salam, in Proceedings of the
8th Nobel Symposium, Stockholm, 1968, edited by N. Svartholm (Almquist and
Wiksells, Stockholm, 1968), p. 367; H. D. Politzer, Phys. Rep. {\bf 14}, 129 (1974.)

\bibitem{higgs}
P. W. Higgs, Phys. Lett. {\bf 12}, 132 (1964); Phys. Rev. Lett. {\bf
13}, 508 (1964); Phys. Rev. {\bf 145}, 1156 (1966); F. Englert and
R. Brout, Phys. Rev. Lett. {\bf 13}, 321 (1964); G. S. Guralnik, C.
R. Hagen and T. W. B. Kibble, Phys. Rev. Lett. {\bf 13}, 585 (1964);
T. W. B. Kibble, Phys. Rev. {\bf 155}, 1554 (1967).

\bibitem{higgs1}
G. Aad \textit{et al}. (ATLAS Collaboration), Phys. Lett. B {\bf 716}, 1 (2012).

\bibitem{higgs2}
S. Chatrchyan \textit{et al}. (CMS Collaboration), Phys. Lett. B {\bf 716}, 30 (2012).

\bibitem{ATLASnew1}
G. Aad \textit{et al}. (ATLAS Collaboration), Phys. Rev. D {\bf 90}, 112015 (2014).

\bibitem{ATLASnew2}
G. Aad \textit{et al}. (ATLAS Collaboration), Phys. Rev. D {\bf 91}, 012006 (2015).

\bibitem{ATLASnew3}
G. Aad \textit{et al}. (ATLAS Collaboration), arXiv:1412.2641 [hep-ex] (2014).

\bibitem{CMSnew1}
V. Khachatryan \textit{et al}. (CMS Collaboration), arXiv:1411.3441 [hep-ex] (2014).

\bibitem{CMSnew2}
V. Khachatryan \textit{et al}. (CMS Collaboration), arXiv:1412.8662 [hep-ex] (2014).

\bibitem{loop-induced1}
G. Cacciapaglia, A. Deandrea, G. D. L. Rochelle and J-B. Flament, J. High Energy Phys. 03 ({\bf 2013}) 029.

\bibitem{loop-induced2}
J.-J. Cao, L. Wu, P.-W. Wu, and J.-M. Yang, J. High Energy Phys. 09 ({\bf 2013}) 043;
C.-C. Han, N. Liu, L. Wu, J.-M. Yang and Y. Zhang, Eur. Phys. J. C {\bf 73}, 2664 (2013).

\bibitem{song2009} S. Mao, W.-G. Ma, R.-Y. Zhang, L. Guo, S.-M. Wang and L. Han,
Phys. Rev. D {\bf 79}, 054016 (2009).

\bibitem{Mao:2013dxa}
  M. Song, N. Wan, G. Li, W.-G. Ma, R.-Y. Zhang, L. Guo, Y.-J. Zhou and J.-Y. Guo,
  Phys. Rev. D {\bf 88}, 076002 (2013).

\bibitem{Liu:2013cla}
  N. Liu, J. Ren, L. Wu, P. Wu and J.-M. Yang,
  J. High Energy Phys. 04 ({\bf 2014}) 189.

\bibitem{HZgamma-anomaly}
M. C. Gonzalez-Garcia, Int. J. Mod. Phys. A {\bf 14}, 3121 (1999);
M. Dubinin, H. J. Schreiber and A. Vologdin, Eur. Phys. J. C {\bf 30}, 337 (2003);
S. D. Rindani and P. Sharma, Phys. Lett. B {\bf 693}, 134 (2010);
A. Gutierrez-Rodriguez, J. Montano, and M. A. Perez, J. Phys. G {\bf 38}, 095003 (2011);
M. B. Einhorn and J. Wudka, Nucl. Phys. {\bf B877}, 792 (2013);
H. Cai, J. High Energy Phys. 04 ({\bf 2014}) 052;

\bibitem{HVV@ee}
M. Baillargeon, F. Boudjema, F. Cuypers, E. Gabrielli, and B. Mele, Nucl. Phys. {\bf B424}, 343 (1994).

\bibitem{fey}
T. Hahn, Comput. Phys. Commun. {\bf 140}, 418 (2001).

\bibitem{formloop}
T. Hahn and M. Perez-Victoria, Comput. Phys. Commun. {\bf 118},
153 (1999).

\bibitem{Passarino}
G. Passarino and M.J.G. Veltman, Nucl. Phys. {\bf B160} 151 (1979).

\bibitem{denner2}
A. Denner and S. Dittmaier, Nucl. Phys. {\bf B734}, 62 (2006).

\bibitem{denner3}
A. Denner, Fortsch. Phys. {\bf 41}, 307 (1993).

\bibitem{ff}
G. J. van Oldenborgh, NIKHEF-H/90-15.

\bibitem{IRDV}
R. K. Ellis and G. Zanderighi,
J. High Energy Phys. 02, ({\bf 2008}) 002.

\bibitem{OneTwoThree}
G.'t Hooft and M. Veltman, Nucl. Phys. {\bf B153}, 365 (1979).

\bibitem{Four}
A. Denner, U. Nierste and R. Scharf, Nucl. Phys. {\bf B367} 637 (1991).

\bibitem{Five}
A. Denner and S. Dittmaier, Nucl. Phys. {\bf B658} 175 (2003).

\bibitem{KLN}
T. Kinoshita, J. Math. Phys. (N.Y.) {\bf 3}, 650 (1962); T. D. Lee and M. Nauenberg, Phys. Rev. {\bf 133},
B1549 (1964).

\bibitem{TCPSS} B. W. Harris and J. F. Owens, Phys. Rev. {\bf D65}
 094032 (2002).

\bibitem{MG5NLO}
J. Alwall, M. Herquet, F. Maltoni, O. Mattelaer, T. Stelzer, JHEP 1106 {\bf 2011} 128.

\bibitem{PDG2012}
J. Beringer \textit{et al}. (Particle Data Group), Phys. Rev. D {\bf 86}, 010001 (2012).

\bibitem{Frixione98}
S. Frixione, Phys. Lett. B {\bf 429}, 369 (1998).

\bibitem{Hdecay}
A. Djouadi, J. Kalinowski, and M. Spira, Comput. Phys. Commun. 108, 56 (1998).

\end{thebibliography}
\end{document}